\documentstyle[aps,prb,preprint,epsfig]{revtex}
\begin{document}

\bibliographystyle{../prsty}

\narrowtext

\title{Resonance in a Tomonaga-Luttinger liquid}
\author{In\`es Safi}
\address{Service de Physique de l'\'Etat Condens\'e, Centre d'\'Etudes de Saclay\\ 91191 Gif-sur-Yvette, France\\}

\maketitle
\begin{abstract}
We study a homogeneous Tomonaga-Luttinger liquid with backscattering potential. A perturbative computation of the conductance at and near resonance is given. We find that the backscattering of one electron dominates that of two electrons for an interaction parameter $K\geq 1/3$ and that the resonance point depends on temperature. Our results may be relevant for recent experiments on shot-noise in FQHE, where the charge $1/3$ and not $2*1/3$ is measured on resonance. 
\end{abstract}
\draft 


One--dimensional electron systems exhibit unique electronic properties described generically by the so called Tomonaga-Luttinger liquid (TLL) \cite{bosonisation}, characterized by phonon-like gapless modes of elementary excitations. One of the typical feature which can differentiate it from the Fermi liquid is the algebraic decay of the correlation function controlled by an interaction parameter. An important consequence, predicted since the seventies\cite{loiT}, is the power law dependence of the conductance when backscattering is present.\cite{loiT,apel_rice,giamarchi_loc,kane_furusaki} Since then a great effort has been devoted to fabricate truly one-dimensional systems, using improvements in nanolithography. It is now possible to obtain ballistic quantum wires\cite{tarucha} up to $10\mu m$. 
Remarkable wires were fabricated more recently;\cite{yacoby_second} the TLL theory could not fully explain the pefect observed plateaus.\\
Another particularly promising system to test the TLL features is the Fractional Quantum Hall effect (FQHE). Edge states can be described by a chiral TLL.\cite{wen} The backscattering by impurities is suppressed, while backscattering between two edges in a Hall bar can be induced in a controllable way using a quantum point contact. Another advantage is that one can tune the gate voltage to achieve a destructive interference for backscattering, thus reaching a resonance. The conductance close to resonance was measured in Ref.\onlinecite{milliken}, and more recently, the tunneling conductance of electrons in a FQHE edge state showed the predicted power law.\cite{chang}\\
Our aim is to study in more detail the behavior at and close to resonance. Our results seem to be relevant for recent experiments on shot noise in FQHE.\cite{saminad_heiblum}

Let us recall briefly the bosonization following Haldane,\cite{haldane_bosons} restricting ourselves to spinless electrons for simplicity. In one dimension, the long-wave part of the electronic density is written as the derivative of a field $\widetilde{\Phi}$, $\rho=-\partial_x\widetilde{\Phi}/\pi$. Thus $\widetilde{\Phi}$ jumps by $\pi$ at each electron location. In order to take into account the discreteness of the charges, the total density is expressed as
\begin{equation}\label{representation}
\rho(x)=-\frac 1\pi\partial_x\widetilde{\Phi}(x)\sum_{m=-\infty}^{+\infty }e^{2im\widetilde{\Phi} \left(x\right)}.
\end{equation} 
One can separate the mean density in $\widetilde{\Phi}=\Phi-k_Fx$. 
As long as long-wavelength properties are concerned, the Hamiltonian can be reduced to
\begin{equation}\label{H}
H=\int_{-L}^L\frac{dx}{2\pi }\left[ uK\Pi^2+\frac
uK(\partial _x\Phi )^2\right],
\end{equation}
where $\Pi$ is the canonical conjugate momentum to $\Phi$: $\left[\Phi(x),\Pi(y)\right]=i\pi\delta(x-y)$ and
$L$ the wire length. $u$ and $K$ are the only parameters needed to describe short-range interactions, and depend on the underlying microscopic model. In the absence of interactions, $K=1$, whereas $K<1$ ($K>1)$ for repulsive (attractive) interactions. \\
$K$ controls the algebraic decay of the density-density correlation functions. For repulsive interactions, the wire develops charge density fluctuations which are pinned by impurities. The conductance of a wire longer than the localization length is easily suppressed.\cite{giamarchi_loc} But it is now possible to fabricate wires much shorter than the mean free path so that they are still conducting; thus transport can include only a few scattering events. The role of one or two local barriers was studied in detail in Ref.\onlinecite{kane_furusaki}. The renormalization group equations indicate that a barrier scales to infinity (zero) for repulsive (attractive) interactions, unless the temperature or the wire length limits this scaling.\\
 Here, we focus on the perturbative conductance for a non-random extended potential configuration. An extended potential renormalizes the interactions,\cite{giamarchi_loc} but this effect can be ignored for weak enough impurities. The renormalization equations will be studied elsewhere.\\
Consider any potential $V(x)$, which couples to the density through the term 
\begin{equation}\label{nouvelle}
H_{imp}=\int \rho(x)V(x)=\sum_{m=-\infty,m\neq 0 }^{+\infty }\int dx \frac {V'(x) }{2i\pi m}e^{2im\widetilde{\Phi}(x)},
\end{equation}
where we have used (\ref{representation}) and performed an integration by parts. 
\label{backscattering}
It is worth making the following remark. A uniform potential has no effect, which is well realized in Eq. (\ref{nouvelle}) since $V'=0$. Had we dropped the terms
\begin{equation}\label{mathcalA}
{\mathcal{A}}_m=-\partial_x\Phi\cos\left[2m\widetilde{\Phi}(x)\right]
\end{equation}  
for $m\neq 0$ in Eq. (\ref{representation}), as is often done in the literature, a uniform potential would give $H_{imp}=V\int \cos 2m\widetilde{\Phi}$ and therefore a notable reduction of the conductance.\\
 Let us now compute the conductance. The conductance of the clean wire is equal to $e^2/h$, a result that can be shown by simulating the reservoirs by the flux they inject in measuring leads.\cite{ines,ines_bis} Using Kubo formula gives the same result\cite{quantum} as long as the reservoirs impose an electro-chemical potential following their electrostatic potential, independently of the current.\cite{ines_bis} In the presence of impurities, we can still use Kubo formula as far as the linear stationary limit is concerned; one needs not to know the profile of the electric field.\cite{ines_bis} We restrict ourselves to high temperatures $T>T_L=u/\pi L$, and consider impurities situated far from the contacts compared to the thermal length $L_T= u/{\pi T}$ so that we can ignore finite size effects.\cite{these_annales}
The perturbative expression for the conductance reads\cite{these_annales}
\begin{equation}\label{gimp}
g_{imp}=\frac{e^2}h\left(1-\mathcal{R}\right),
\end{equation}
where $\mathcal{R}$ designates
\begin{equation} \label{RsommeRm}
{\mathcal{R}}=\sum_{m=1}^\infty {\mathcal{R}}_{m},
\end{equation}
and ${\mathcal{R}}_{m}$ is the contribution from backscattering of $m$ electrons. Up to non-universal prefactors depending on the short distance behavior, we find 
\begin{eqnarray} \label{eqRmnew}
{\mathcal{R}}_m& =&-\int\-\-\-\int dxdy\underline{T}^{2(m^2K-1)}C_{m^2K}\left( \frac{x-y}{L_T}\right)\\\nonumber&& \frac{V'(x)V'(y)}{(k_Fu)^2}\cos 2mk_F(x-y),
\end{eqnarray}
 where
\begin{equation}
C_K\left( v\right)=B\left( \frac 12,K\right)F\left( K,K,\frac 12+K;-\sinh ^2v\right),  \label{CKvtexte}
\end{equation}
$B$ is the Beta function and $F$ is the Hypergeometric function with one variable. $C_K$ is even, and its odd derivatives vanish at $v=0$. 
 For $v\gg 1$:
\begin{equation}\label{decay}
C_K(v\gg 1)=2^Kve^{-2Kv}.
\end{equation} 
In Eq. (\ref{eqRmnew}), $\underline{T}$ designates
\begin{equation}\label{Tscale}
\underline{T}=\pi T/u k_c,
\end{equation}
where $k_c$ is a cutoff different from $k_F$. It is worth noting that a power law dependence on $T$ of $\mathcal{R}$ [Eq. (\ref{eqRmnew})] holds only if the total extension of the potential is much less than $L_T$. In this case, any $r$ in the integrand satisfies $r\ll L_T$, and we can expand $C_K(r/L_T)$ in powers of $r/L_T$. The complete formal power expansion of ${\mathcal{R}}_m$, Eq. (\ref{eqRmnew}) in temperature or in $L_T$ is 
\begin{equation}\label{devR}
{\mathcal{R}}_m=\left(\frac{k_c}{k_F}\right)^2\underline{T}^{2(m^2K-1)}\sum_{n=0}^{\infty}\frac{(-1)^nC_{K}^{(2n)}(0){W^{(2n)}(2mk_F)}}{2n!L_T^{2n}},
\end{equation}
where $W^{(2n)}$ designates the $2n$th derivative of
\begin{equation}\label{W}
W(k)=\left(\frac{k}{k_c}\right)^2\vert V(k)\vert ^2,
\end{equation}
and $V(k)$ is the Fourier transform of $V(x)/u$. Let us comment the role of the operators ${\mathcal{A}}_m$, Eq. (\ref{mathcalA}). Had we dropped them, we would have to substitute $\alpha V'(x)\rightarrow -imV(x)$ in Eq. (\ref{eqRmnew}), thus $W(k)$ Eq. (\ref{W}) is replaced by $\vert  V(k)\vert^2$. This  would lead to similar results as Refs.\onlinecite{these_annales,geller_loss,chamon_freed}. The operators ${\mathcal{A}}_m$ change the prefactors for $n\geq 2$ in Eq. (\ref{devR}) (see the remark above).\\ In the limit of a local barrier, $V(x)=Vu\delta(x)$. In this case, $V(2k_F)=V$ independent of $k_F$ and the development (\ref{devR}) stops at $n=2$, so that Eq. (\ref{RsommeRm}) for $m=1$ yields
\begin{eqnarray}\label{devRlocal}
{\mathcal{R}}_1&=&V^2\left[B(\frac 1 2,K)\underline{T}^{2(K-1)}+KB(\frac 1 2,K+1)\underline{T}^{2K}\right],\\\nonumber
\end{eqnarray}
while ${\mathcal{R}}_2\sim V^2\underline{T}^{2(4K-1)}$ if one retains only $n=0$. As in Refs.\onlinecite{kane_furusaki,furusaki_deux_barrieres}, the dominant contribution is ${\mathcal{R}}\sim V^2\underline{T}^{2(K-1)}$; for $K>1$, $\mathcal{R}$ decreases with temperature saturating at $T<T_L=u/L\pi$. For $K<1$, $\mathcal{R}$ increases  up to its value at $T_L$, so that we must have $\underline{T}_L^{K-1}\vert V\vert\ll 1$ for the perturbation to be valid in all temperature range. \\
In order to obtain the second dominant term, we have to compare the power $\underline{T}^{2K}$ to ${\mathcal{R}}_2$. Their ratio $\underline{T}^{2(-3K+1)}$ equals unity for $K=1/3$, is $\gg 1$ for $K>1/3$ and $\ll 1$ for $K<1/3$. Thus the second dominant part does not come from the backscattering of two electrons when $K>1/3$, {\em but still from that of one electron}. For such a local barrier, this is due to the fact that ${\mathcal{A}}_1$, Eq. (\ref{mathcalA}) has a dimension $K+1$, and is a more relevant operator than $e^{4i\Phi}$ for $K>1/3$. \\
The same discussion holds for a more extended potential. For repulsive interactions, the conductance can vanish at low enough temperature or for strong barrier. But $\vert V(2k_F)\vert ^2\underline{T}^{2(K-1)}$ cancels out if $V(2k_F)$ vanishes at values of $k_F$ one can reach by tuning the gate voltage, thus recovering the ballistic conductance at zero temperature. This is the resonance situation we are going to focus on.
First, let us remark that in order to achieve resonance for a fixed potential configuration, the temperature has to be much less then the inverse of its total extension; for instance, if two impurities are separated by $a\gg L_T$, eqs.(\ref{decay},\ref{eqRmnew}) show that the contribution from $r=a$ in Eq. (\ref{eqRmnew}) is $ V^2T^{2m^2K-1}a\exp({-2Km^2a/L_T}).$ Thus they act like two separate impurities, and it is difficult to cancel simultaneously their individual contributions (\ref{devRlocal}). Then we restrict ourselves to a potential whose total extension is less than $L_T$, in which case one can retain only the first terms in Eq. (\ref{devRlocal}). In the following, we denote $\delta= \vert V(2k_F)\vert$. We will consider first the situation on resonance where $\delta=0$, then close to resonance where $\delta\ll 1$.\\
 A value of $2k_F$ at which $\delta=0$ is denoted by $k_0$.  Ignoring terms in Eq. (\ref{devR}) for $m=1$ which can dominate ${\mathcal{R}}_2$, Eq. (\ref{RsommeRm}) becomes
\begin{equation}\label{onresonance}
{\mathcal{R}}(\delta=0)= k_0^2\vert V'(k_0)\vert^2\underline{T}^{2K}+\vert V(2k_0)\vert^2 \underline{T}^{2(4K-1)}.
\end{equation}
 Note that the second term is irrelevant for $K>1/4$: this ensures perfect resonance because ${\mathcal{R}}(\delta=0)=0$ at zero temperature.\cite{kane_furusaki,furusaki_deux_barrieres} As for a local barrier, let us compare the first and second terms emanating respectively from the $2k_F$ and $4k_F$ backscattering.
For $K>1/3$, the first term dominates the second. For $K=1/3$, both terms are in $\underline{T}^{2/3}$, and their relative effect is determined by the ratio 
\begin{equation}\label{lambda}
\lambda=\left\vert\frac{ k_0 V'(k_0)}{V(2k_0)}\right\vert. 
\end{equation}
For $K<1/3$, the $4k_F$ contribution dominates. But the situation is reversed whenever $\lambda \gg 1$ and for temperatures greater than

\begin{equation}
\label{Tstar}
\underline{T}^*=\lambda ^{1/(-1+3K)}.
\end{equation} 
In this case, one recovers the power $T^{2K}$. \\
 
\begin{figure}[htb]
\begin{center}
\epsfig{file=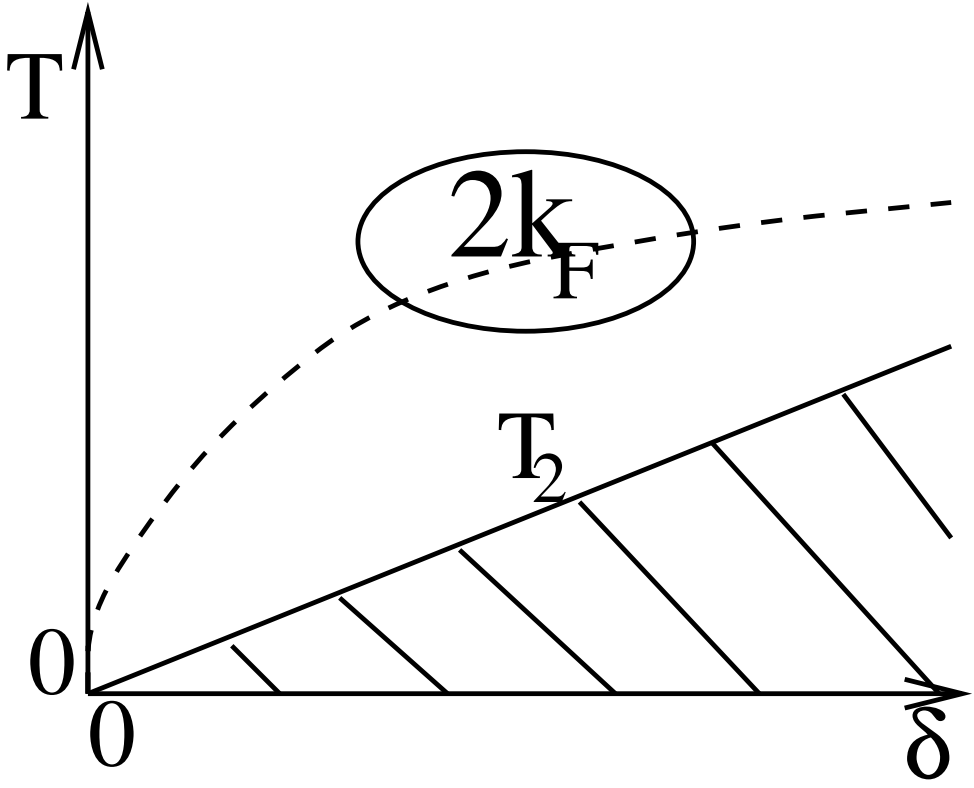,width=13cm}
\caption{$K>1/3$: dominant reduction to the conductance close to resonance in the plane $T$, $\delta\ll 1$. $T$ is scaled by $uk_c$. The hatched region where $X^2$ Eq. (\ref{X}) dominates is limited by the line $T_2=\delta/\vert k_0V'(k_0)\vert$. Above $T_2$, the $2k_F$ backscattering contribution on resonance $T^{2K}$ dominates. The dashed curve indicates the limit one would infer from the previous works, $T_4$ is given by Eq. (\ref{T4}), and above which the $4k_F$ would dominate. The $2k_F$ backscattering dominates everywhere.}

\label{f:Kmore}
\end{center}
\end{figure}

\begin{figure}[ptb]
\begin{center}
\epsfig{file=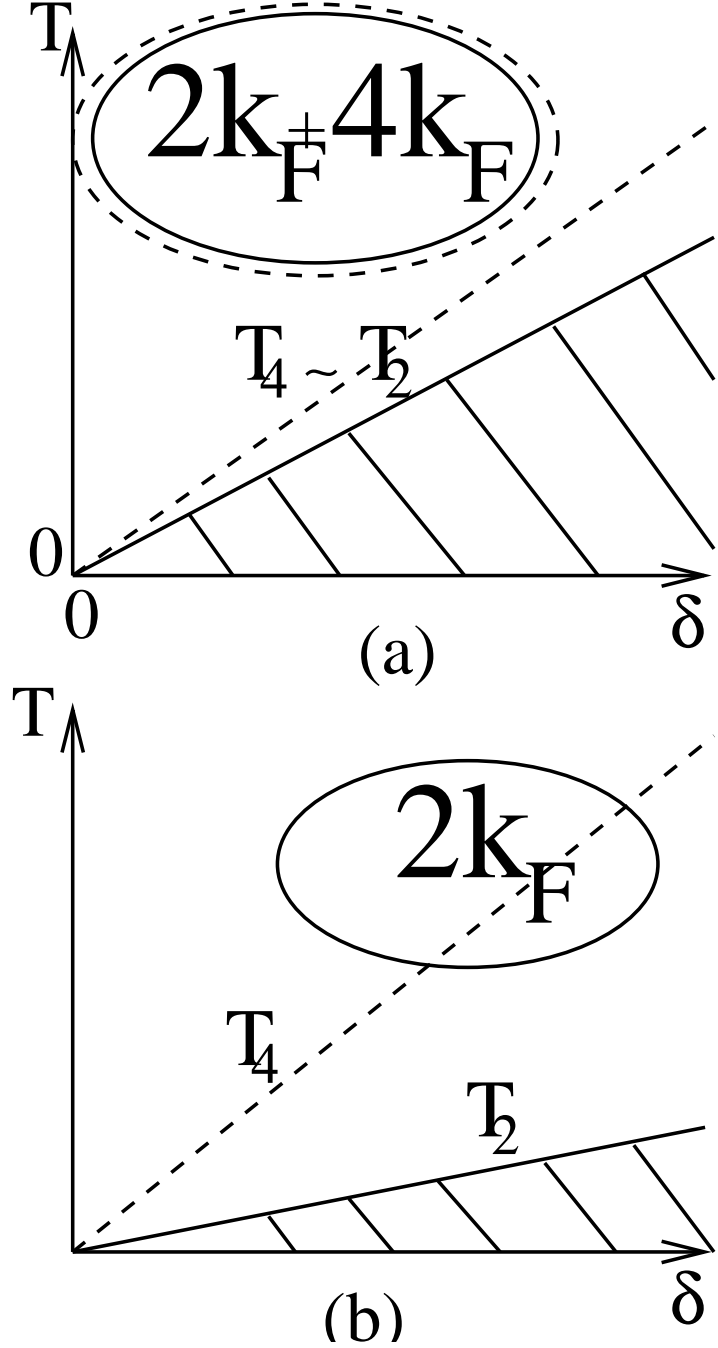,height=18cm}
\caption{$K=1/3$: $T_4$, Eq. (\ref{T4}) is now linear in  $\delta$. Now $T_4/T_2=\lambda$, Eq. (\ref{lambda}). In Fig. (a), $\lambda\sim1$: both contributions of the $2k_F$ and $4k_F$ backscattering dominate above $T_4\sim T_2$ and are in $T^{2/3}$. For $\lambda>>1$, fig.(b), $T_4\gg T_2$; the $2k_F$ backscattering dominates all over the region $T\gg T_2$, and the domain where $X^2$ dominates is shrinked.}
\label{f:Kequal}
\end{center}
\end{figure}

\begin{figure}[ptb]
\begin{center}
\epsfig{file=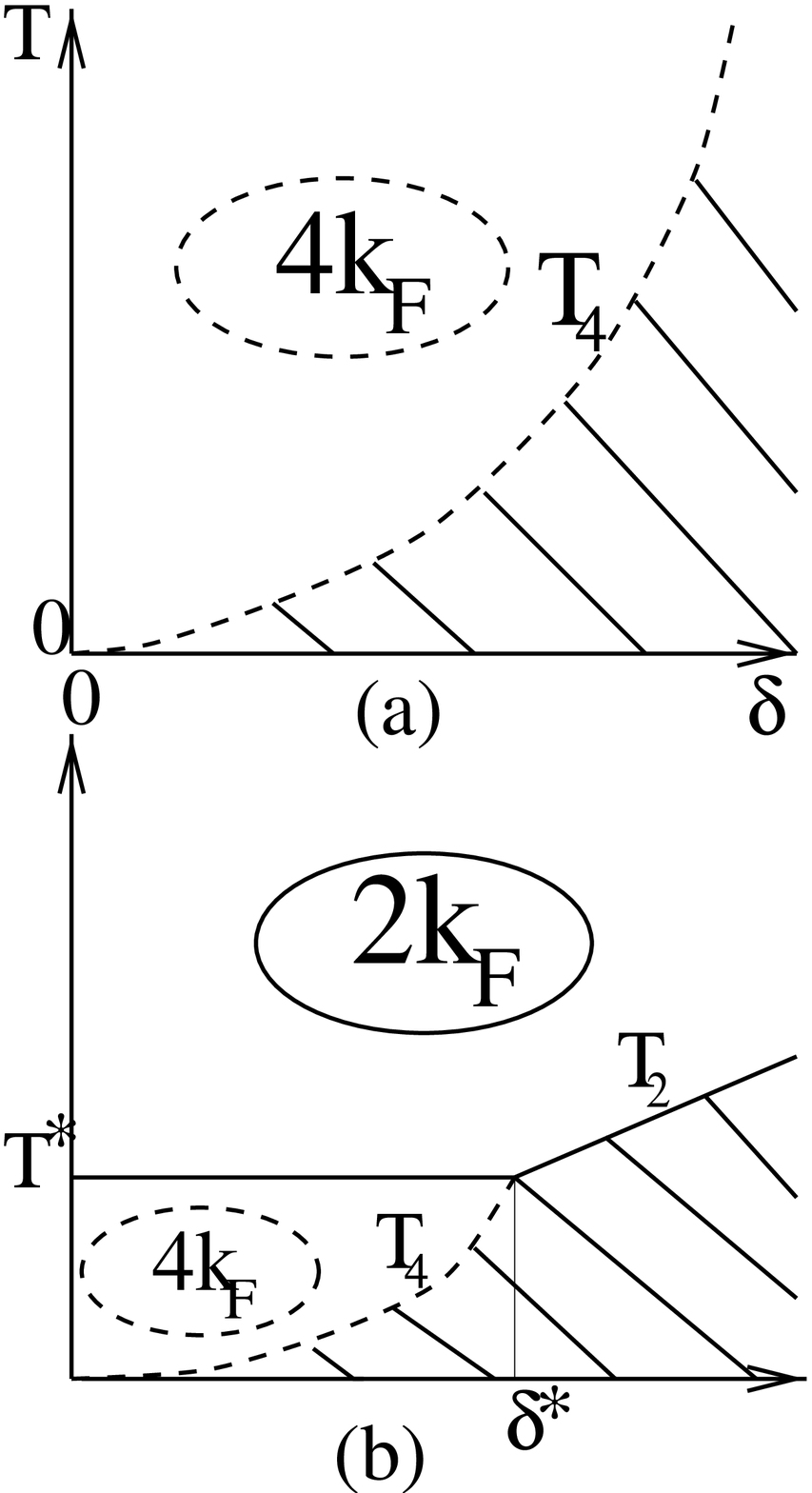,height=18cm}
\caption{$K<1/3$. In (a), $\lambda\sim 1$: the $4k_F$ backscattering dominates above $T_4$. In (b), $\lambda \gg 1$. $4k_F$ dominates only in between $T_4$ and $T^*$ [Eq. (\ref{Tstar})] and for $\delta\ll \delta^*=T^*\vert k_0V'(k_0)\vert$, while the $2k_F$ backscattering dominates everywhere above $\max(T^*,T_2)$.}
\end{center}
\label{f:Kless}
\end{figure}

We consider now the conductance close to resonance. Then ${\mathcal{R}}(\delta)={\mathcal{R}}(0)+X^2$, where ${\mathcal{R}}(0)$ is given by Eq. (\ref{onresonance}) and
\begin{equation}\label{X}
X=\delta\underline{T}^{K-1}.
\end{equation} 
In Refs.\onlinecite{kane_furusaki,fendley}, it was claimed, using the scaling trajectory, that the conductance close to resonance is a universal function of $X$, and that it behaves as $1-X^2$ for small $X$. We will determine in a more precise way the domain in the plane $\delta,T$ where this holds. In this domain, $X^2$ dominates the resonance value ${\mathcal{R}}(0)$ (\ref{onresonance}). Outside this domain, ${\mathcal{R}}(0)$ dominates $X^2$ and one is back to the on-resonance situation above where we have compared the $2k_F$ to the $4k_F$ backscattering contribution. \\
Now $X^2$ dominates the first term in ${\mathcal{R}}(0)$ for temperatures less than
\begin{equation}\label{T2}
 \underline{T}_2=\frac{\delta}{\vert k_0 V'(k_0)\vert},
\end{equation}
while $X^2$ dominates the second term in Eq. (\ref{onresonance}) at temperatures less than
\begin{equation}
\label{T4}
\underline{T}_4=\left(\frac{\delta}{V(2k_0)}\right)^{1/3K}.
\end{equation}
Remember that $\underline{T}=T/uk_c$. 
The three cases: $K>1/3$, $K=1/3$ and $K<1/3$, are illustrated by figures \ref{f:Kmore}, \ref{f:Kequal} and \ref{f:Kless}. The hatched parts in these figures correspond to the dominance of $X^2$. The continuous linear curves designate $T_2$ Eq. (\ref{T2}), while the dashed curves designate $T_4$, Eq. (\ref{T4}). 

Consider now the special case $V(x)=uV\left[\delta(x-a/2)+\delta(x+a/2)\right]$, {\it i.e.} two symmetric local barriers distant by $a$. As noted before, one can reach resonance if $a\ll L_T$. The function $\vert V(k)\vert ^2=2V^2(1+\cos ka)$ vanishes at $k_0=(2n+1)\pi/a$ for any $n$, and Eq. (\ref{lambda}) becomes $\lambda=a k_0.$
 If the barriers are much more distant than the mean electron spacing, we have $\lambda\gg 1$. In particular, for $K=1/3$, the backscattering of one electron still dominates that of two electrons. Note also that Eq. (\ref{T2}) becomes $\underline{T}_2=\delta/(\lambda V)\ll\delta/V\ll 1$; this {\em shrinks considerably the temperature range} where the conductance depends only on $X^2$, Eq. (\ref{X}). 

In a real experiment, the resonance is reached by tuning the gate voltage that modifies the density or $k_F$ in a way depending on capacitive effects. There is another ingredient which has to be taken into account: the interaction parameter $K$ depends on $k_F$, thus changes with gate voltage.\cite{remark5} This adds a linear term in $\delta$ to $\mathcal{R}$. The subdominant term in ${\mathcal{R}}_1$ gives the dominant shift of the resonance momentum $k_0$, equal to  
\begin{equation}\label{translationK}
\frac{dK}{dk_F}k_c^2\underline{T}^{2}\log \underline{T}.
\end{equation} Thus the resonance point varies with temperature. It is interesting to note that this change does not depend on the backscattering potential $V$. A rough estimate of $K$ is $K=\left(1+{U}/{2E_F}\right)^{-1/2}$
 where $U$ is the screened Coulomb interaction, so that: $
2k_F{dK}/{d k_F}=K(K^2-1)$. But, in general, it is not an easy task to derive a microscopic expression for $K$, which requires to know the way interactions are screened by surrounding gates.\\  Recall that the backscattering potential has to be symmetric\cite{kane_furusaki,furusaki_deux_barrieres} or antisymmetric so that $V(2k_F)$ is purely real or imaginary, thus can be suppressed by tuning one parameter.

Indeed, resonance is much easier to realize experimentally in edge states than in quantum wires; we focus on the FQHE in the following. 
The weak backscattering between two opposite edges can be treated in an analogous way to spinless electrons, $K$ being replaced by the filling factor $\nu$. Without backscattering, the conductance has now to be renormalized by $\nu$. With backscattering, the first harmonic in Eq. (\ref{representation}),  $m=1$  emerges from the product of the fermion operator on the right edge by that on the left edge. It is not clear how one can give a microscopic justification for the other terms. Normally, terms of the type $\cos(2m\Phi)$ are generated by the renormalization procedure; the $4k_F$ backscattering has a prefactor $\vert V_4\vert\sim V^2$. This can increase the ratio $\lambda$, Eq. (\ref{lambda}), and therefore makes the $4k_F$ backscattering contribution less important than $2k_F$ at $\nu=1/3$. Besides, ${\mathcal{A}}_m$, Eq. (\ref{mathcalA}) has not to be included in Eq. (\ref{representation}). In this case, as quoted previously, the expansion (\ref{devR}) is still valid, but with $W(k)$, Eq. (\ref{W}), replaced by $\vert  V(k)\vert^2$. Remarkably, this does not affect the prefactors close to resonance where the expression (\ref{onresonance}) plus $X^2$ Eq. (\ref{X}) still holds.\cite{remark6} One has exactly the same results as those illustrated in Figs. (\ref{f:Kmore},\ref{f:Kequal},\ref{f:Kless}) corresponding respectively to $\nu>1/3$, $\nu=1/3$ or $\nu<1/3$.\\ The interpretation of $\nu$ is different from that of $K$. For instance, due to incompressibility, one can maintain $\nu$ fixed while gate voltage $V_G$ is varied to draw the two edge states closer. Thus one does not expect a translation of the point resonance as in Eq. (\ref{translationK}).
 But more generally, the point resonance can be shifted if $\mathcal{R}$ contains any function of $V_G$ with a non vanishing first derivative at resonance. An additional remark is that $V_G$ can modify the profile of $V$ itself, and that $k_F$ can now depend on the separation between the two edges, thus on $x$.  \\
Let us now discuss recent remarkable experiments\cite{saminad_heiblum} allowing a direct observation of the fractional charge $1/3$. The shot noise is measured for two edge states at filling factor $1/3$ drawn closer by applying locally a gate voltage. At resonance, it was predicted that the observed charge should be multiplied by $2$ because the backscattering is then dominated by that of two quasiparticles.\cite{kane_fisher_noise} Nevertheless, our study has shown that at $\nu=1/3$, the backscattering of one quasiparticle yields a comparable contribution to the two quasiparticles backscattering and can even dominate it. This may be an explanation of the observed charge $1/3$ and not a pair of quasiparticles $2*1/3$ on resonance in Ref.\onlinecite{saminad_heiblum}.\\

To conclude, we have computed perturbatively the conductance of a Tomonoga-Luttinger liquid in the presence of extended backscattering potential, retaining supplementary terms whose role was ignored in previous works. Out of resonance, we recover the same dominant correction as that found previously.\cite{kane_furusaki}
At resonance, we have showed that the backscattering of one electron still dominates that of two electrons as far as $K>1/3$, giving a power $T^{2K}$. For $K=1/3$, both processes are degenerate, but the $2k_F$ backscattering can still dominate. At $K<1/3$, the competition depends on the temperature and on the prefactors. The new power $T^{2K}$ has consequences close to resonance: the domain over which the conductance depends only on the combination $V^2T^{2(K-1)}$ is shrinked; this effect is enhanced, for instance, for two barriers well separated compared to $\lambda_F$. Furthermore, the dependence of $K$ on gate voltage induces a temperature dependent shift of the point resonance. Finally, we have discussed the relevance of our results for recent experiments on FQHE\cite{saminad_heiblum} at filling factor $1/3$.\\

 The author is grateful to T. Giamarchi, D. C. Glattli, T. Martin, L. Saminadayar and H. J. Schulz for stimulating discussions.

\begin{thebibliography}{10}
\bibitem{remark6}
Although the subdominant terms of ${\mathcal{R}}_1$ were implicitly present in
  some of the previous
  works\cite{furusaki_deux_barrieres,these_annales,chamon_freed}, their role on
  and close to resonance has not been investigated.
\end{thebibliography}

\end{document}